\documentclass[letterpaper,twocolumn,prl,aps,superscriptaddress,amsmath,amssymb,floatfix]{revtex4-2}
\usepackage{mathptmx}
\usepackage[utf8]{inputenc}
\usepackage[T1]{fontenc}
\usepackage{color}
\usepackage{float}
\usepackage{amsmath}
\usepackage{amssymb}
\usepackage{graphicx}
\usepackage{esint}
\usepackage[unicode=true,
 bookmarks=true,bookmarksnumbered=false,bookmarksopen=false,
 breaklinks=false,pdfborder={0 0 1},backref=false,colorlinks=true]
 {hyperref}
\hypersetup{
 linkcolor=magenta,urlcolor=blue,citecolor=blue,pdfstartview={FitH},hyperfootnotes=false}

\makeatletter

\pdfpageheight\paperheight
\pdfpagewidth\paperwidth

\usepackage{textcomp}
\usepackage{epstopdf}

\usepackage{amsfonts}

\pdfpageheight\paperheight
\pdfpagewidth\paperwidth

\@ifundefined{textcolor}{}{%
 \definecolor{BLACK}{gray}{0}
 \definecolor{WHITE}{gray}{1}
 \definecolor{RED}{rgb}{1,0,0}
 \definecolor{GREEN}{rgb}{0,1,0}
 \definecolor{BLUE}{rgb}{0,0,1}
 \definecolor{CYAN}{cmyk}{1,0,0,0}
 \definecolor{MAGENTA}{cmyk}{0,1,0,0}
 \definecolor{YELLOW}{cmyk}{0,0,1,0}
}

\usepackage{xcolor}\usepackage{soul}
\newcommand{\ket}[1]{\ensuremath{\left|#1\right\rangle}}

\definecolor{blue}{rgb}{0,0,1}
\definecolor{red}{rgb}{1,0,0}
\definecolor{green}{rgb}{0,1,0}

\usepackage{soul}

\makeatother
\begin{document}
\title{Lensing and enhanced single atom detection via a single-pixel nanostructure}

\author{Ling-Xiao Wang}
\thanks{These authors have contributed equally to this work.}
\affiliation{Laboratory of Quantum Information, University of Science and
Technology of China, Hefei 230026, China}
\affiliation{Anhui Province Key Laboratory of Quantum Network, University of Science and Technology of China, Hefei 230026, China}

\author{Lei Xu}
\thanks{These authors have contributed equally to this work.}
\affiliation{Laboratory of Quantum Information, University of Science and
Technology of China, Hefei 230026, China}
\affiliation{Anhui Province Key Laboratory of Quantum Network, University of Science and Technology of China, Hefei 230026, China}

\author{Ai-Ping Liu}
\affiliation{Institute of Quantum Information and Technology, Nanjing University of Posts and Telecommunications, Nanjing 210003, China}

\author{Guang-Jie Chen}
\affiliation{Laboratory of Quantum Information, University of Science and
Technology of China, Hefei 230026, China}
\affiliation{Fujian Provincial Key Laboratory of Quantum Manipulation and New Energy Materials, College of Physics and Energy, Fujian Normal University, Fuzhou 350117, China}

\author{Yuan-Hao Yang}
\affiliation{Laboratory of Quantum Information, University of Science and
Technology of China, Hefei 230026, China}
\affiliation{Anhui Province Key Laboratory of Quantum Network, University of Science and Technology of China, Hefei 230026, China}

\author{Jia-Qi Wang}
\affiliation{Laboratory of Quantum Information, University of Science and
Technology of China, Hefei 230026, China}
\affiliation{Anhui Province Key Laboratory of Quantum Network, University of Science and Technology of China, Hefei 230026, China}

\author{Xin-Biao Xu}
\affiliation{Laboratory of Quantum Information, University of Science and
Technology of China, Hefei 230026, China}
\affiliation{Anhui Province Key Laboratory of Quantum Network, University of Science and Technology of China, Hefei 230026, China}


\author{Guang-Can Guo}
\affiliation{Laboratory of Quantum Information, University of Science and
Technology of China, Hefei 230026, China}
\affiliation{Anhui Province Key Laboratory of Quantum Network, University of Science and Technology of China, Hefei 230026, China}
\affiliation{CAS Center for Excellence in Quantum Information and Quantum Physics,
University of Science and Technology of China, Hefei 230026, China}
\affiliation{Hefei National Laboratory, University of Science and Technology of China, Hefei 230088, China}

\author{Chang-Ling Zou}
\email{clzou321@ustc.edu.cn}
\affiliation{Laboratory of Quantum Information, University of Science and
Technology of China, Hefei 230026, China}
\affiliation{Anhui Province Key Laboratory of Quantum Network, University of Science and Technology of China, Hefei 230026, China}
\affiliation{CAS Center for Excellence in Quantum Information and Quantum Physics,
University of Science and Technology of China, Hefei 230026, China}
\affiliation{Hefei National Laboratory, University of Science and Technology of China, Hefei 230088, China}

\author{Guo-Yong Xiang}
\email{gyxiang@ustc.edu.cn}
\affiliation{Laboratory of Quantum Information, University of Science and
Technology of China, Hefei 230026, China}
\affiliation{Anhui Province Key Laboratory of Quantum Network, University of Science and Technology of China, Hefei 230026, China}
\affiliation{CAS Center for Excellence in Quantum Information and Quantum Physics,
University of Science and Technology of China, Hefei 230026, China}
\affiliation{Hefei National Laboratory, University of Science and Technology of China, Hefei 230088, China}

\date{\today}

\begin{abstract}
We propose and demonstrate a general mechanism for nanoscale lensing based on the phase gradient imposed by a single nanostructure scattering light in its near-field. We verify this effect using an optical waveguide on a substrate, with single atoms serving as quantum probes that sample the near-field intensity through their fluorescence. This quantum probing technique provides a unique, non-destructive approach to characterizing focused optical fields and reveals a 4-fold enhancement in single atom detection efficiency. This work establishes on-chip nanostructures as a multi-functional quantum optics platform that can efficiently route photons, localize fields, and enhance atom-photon coupling, offering new opportunities for trapping and manipulating single atoms and realizing hybrid nanophotonic-atomic systems for quantum applications.
\end{abstract}

\maketitle

\noindent \textbf{\emph{Introduction.-}} 
Hybrid photonic-atomic systems have emerged as a promising platform for scalable quantum information processing, combining the advantages of nanophotonic circuits for efficient photon manipulation~\cite{wang2020,minzioni2019} and single emitters for quantum nonlinearity~\cite{chang2014,Liu2023,Yang2023} and memory~\cite{specht2011,Corzo2019}. Integrated photonic chips offer a compact, stable, and scalable infrastructure for routing and confining photons at the nanoscale~\cite{Pelucchi2022,Chang2019}, while individual quantum emitters, such as atoms, quantum dots, and color centers, provide the essential single-photon nonlinearity~\cite{Hacker2016,Javadi2015,Bhaskar2017} and quantum memory~\cite{Kalb2015,Waldermann2007} capabilities required for quantum logic operations~\cite{Hacker2016,Grinkemeyer2025} and network nodes~\cite{Andreas2022,Hartung2024,hu2025,Wang2025}. Interfacing these two complementary platforms holds great promise for realizing complex quantum processors and networks with enhanced functionality and scalability. Consequently, significant efforts have been devoted to integrating various single emitters onto photonic chips, including quantum dots, neutral atoms, molecules, and defects in solids~\cite{Javadi2015,Zhou2023,Menon2024,Kim2020,Wan2020}. Among these systems, neutral atoms stand out as excellent candidates due to their indistinguishability, long coherence times, and the ability to deterministically assemble single-atom devices using optical tweezers~\cite{Kaufman2021,Dordevic2021}.

However, efficiently loading and probing single atoms in the near-field of  nanophotonic devices remains an outstanding challenge. On one hand, optical dipole traps, such as free-space optical tweezers, provide a powerful tool for manipulating single atoms with high precision and fidelity~\cite{Seubert2025,Brown2023}. Yet, the presence of nearby nanophotonic structures can significantly alter the trapping potential landscape~\cite{Zhou2024,Thompson2013,Luan2020}, making it difficult to create tightly focused and stable traps compatible with the photonic geometry. On the other hand, detecting single atoms close to photonic structures is complicated by the modification of atomic emission and the collection efficiency imposed by the complex electromagnetic environment~\cite{Menon2024,Kim2019,Meng2020}. The scattered and radiated fields from the atoms can interact strongly with the nanophotonic modes and subwavelength features, leading to a substantial reduction in the atomic fluorescence collected by conventional free-space optics. Overcoming these challenges in atom trapping and detection near nanophotonic interfaces is crucial for harnessing the full potential of hybrid photonic-atomic platforms~\cite{Will2021,Samutpraphoot2020,Han2026,Liu2022}.

\begin{figure*}[hbt]
\centering
\includegraphics[width=1.0\textwidth]{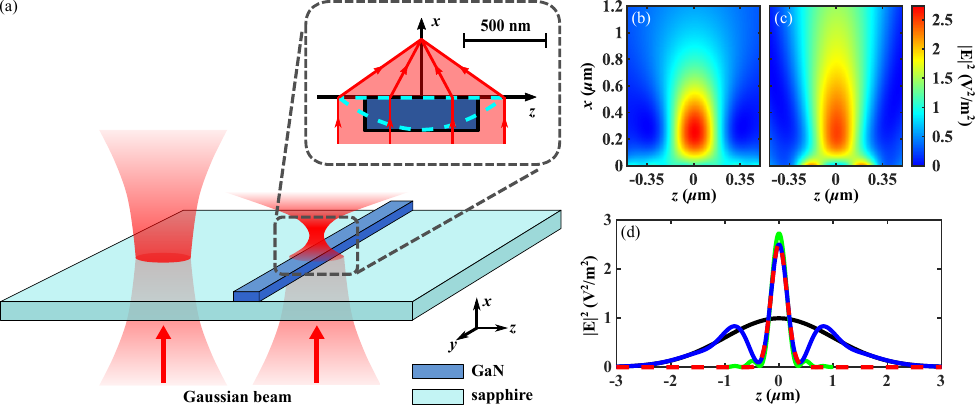}
\caption{(a) Schematic of lensing effect. An incident plane wave passing across a nanostructure (blue rectangle) acquires a spatially varying phase and forms a focal spot in the near-field of the nanostructure. The cyan dashed line outlines the equivalent lens. (b) Calculated evolution of light field in space using the angular spectrum method. (c) Numerical simulated $|\mathbf{E}|^2$ distribution of the total field in the vicinity of a trapezium dielectric waveguide illuminated from below by a Gaussian beam. (d) Green line: the transverse $|\mathbf{E}|^2$ profile at the focal plane calculated by the angular spectrum method. Blue line: the numerical simulated transverse $|\mathbf{E}|^2$ profile at the focal plane. Red dashed line: Gaussian fit of the central peak of the blue line. Black line: the $|\mathbf{E}|^2$ profile of the incident Gaussian beam without a nanolensing effect. Here, the parameters for simulations are $a=700\,\mathrm{nm}$, $t=200\,\mathrm{nm}$, $n=2.36$, and $\lambda=780\,\mathrm{nm}$.}
\label{fig1}
\end{figure*}

In this Letter, we propose and demonstrate a novel approach to enhance the loading and probing of single atoms in nanophotonic devices by harnessing the optical potential of the photonic structure itself. We show that the intrinsic phase gradient associated with the scattering of light by a nanophotonic element can naturally create a nanoscale lensing effect, leading to subwavelength focusing and confinement of optical fields. As a proof-of-principle demonstration, we investigate the nanolensing effect of a waveguide using single atoms as sensitive probes and observe a fourfold enhancement in the single atom detection efficiency. Our work establishes a novel approach for atom-nanophotonic interfaces, where nanophotonic elements serve the dual purpose of focusing light for atom trapping and enhancing light-matter coupling for efficient atom-photon interactions, thereby opening new possibilities for realizing hybrid nanophotonic-atomic systems that exploit nanoscale engineering to control and enhance the quantum dynamics of atoms and photons on a chip.

\smallskip{}
\noindent \textbf{\emph{Principle of lensing effect.-}} 
Figure~\ref{fig1}(a) illustrates the Gaussian beam passing through a photonic chip to interact with single atoms near the chip surface. When there are no photonic structures, the Gaussian beam can pass through the substrate as expected with only partial reflection of input light, while nanostructures would introduce significant scattering loss of the beam and also distort the beam profile. However, we predict that a simple single-pixel nanostructure, or equivalently, a nanostructure that can be effectively treated as a single pixel with all features being negligible compared with the optical wavelength ($\lambda$), can induce focusing of the input beam instead of random distortion. 

As shown in the inset of Fig.~\ref{fig1}(a), a waveguide structure can produce the lens effect, as a waveguide has a higher dielectric constant (blue rectangle) that generates a phase gradient to the central part of the input beam. Considering a simplified 2D model of a pixel (width $a$ and thickness $t$), an approximate quadratic phase profile will be imprinted to the transmitted optical field. Assuming a uniform input optical field, we have a complex amplitude distribution of electric field after passing through the pixel as $E_0(z)=\mathrm{exp}[-i\frac{2\mathrm{\pi}(n-1)t}{\lambda}\frac{2z^2}{a^2}]$, where $n$ is the refractive index of the pixel. Here, due to the finite thickness of the pixel ($t$), the edge of the pixel should not generate a discontinuous phase accumulation. The quadratic approximation is valid to the fourth order ($\mathcal{O}(z^4)$) due to the symmetry. Such a phase accumulation mimics a thin lens, as shown by the cyan dashed line. Then, the resulting field distribution above the pixel can be analytically solved by the angular spectrum method for a plane wave input, as shown in Fig.~\ref{fig1}(b). As expected, a focused Gaussian-like beam is generated with a beam waist of $280\, \mathrm{nm}$ and a focal point about $250\, \mathrm{nm}$ above the pixel. The lens effect is also confirmed by numerical simulations in Fig.~\ref{fig1}(c), while the input Gaussian beam has a waist of $2\,\mathrm{\mu m}$. The electric field intensity ($|\mathbf{E}|^2$) at the focal plane [Fig.~\ref{fig1}(d)] reveals a magnification of input light intensity by $\alpha\approx2.5$, with the profiles fitted by a Gaussian function with a waist of $280 \,\mathrm{nm}$, highlighting the strong localization and confinement of light by the nanolens~\cite{SM}.

\begin{figure}[t]
\centering
\includegraphics[width=1.0\columnwidth]{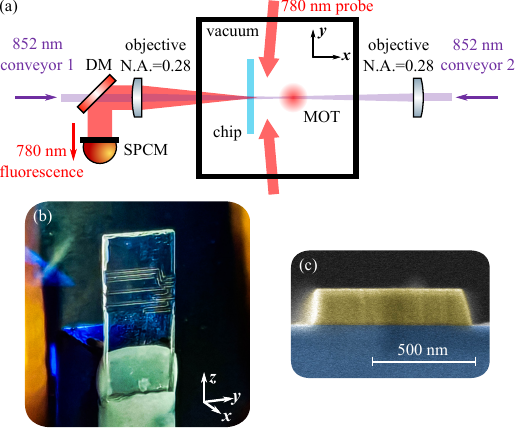}
\caption{(a) Illustration of experimental setup. A fully transparent GaN-on-sapphire chip is fixed in the center of a vacuum cell. Cold $\mathrm{^{87}Rb}$ atoms are prepared by a magneto-optic trap (MOT) about $550\,\mathrm{\mu m}$ from the chip surface and then transported to the chip surface by an optical conveyor belt consisting of two z-direction linearly polarized counterpropagating Gaussian beams. The probe beam grazes on the chip with an incident angle of $85 ^\circ$. The fluorescence of the excited atoms is collected by the same objective used for focusing conveyor 1 beam and separated from conveyor beam by a dichroic mirror (DM). After polarization filtering, atom fluorescence is coupled into a single-mode fiber and detected by a single-photon counting module (SPCM). (b) Photo of the chip in the vacuum, on which the waveguides can be clearly seen. (c)  Scanning electron micrograph (false color) of the waveguide cross section on the chip.}
\label{fig2} 
\end{figure}

\smallskip{}
\noindent \textbf{\emph{Experimental Setup.-}} 
We experimentally demonstrate the nanolensing effect by measuring the enhanced collection of single atoms fluorescence in the near-field of a waveguide, and the corresponding experimental setup is shown in Fig.~\ref{fig2}(a). In the center of a vacuum cell, a fully transparent sapphire chip is fixed [Fig.~\ref{fig2}(b)], with multiple gallium nitride (GaN) waveguides fabricated on the chip\cite{Zheng2022,Xu2025}. Figure~\ref{fig2}(c) illustrates the cross-section of the waveguide, showing the thickness of $t=200\,\mathrm{nm}$ and the width of $a=700\,\mathrm{nm}$. Cold $\mathrm{^{87}Rb}$ atoms are prepared by a magneto-optic trap (MOT), with the center being about $550\,\mathrm{\mu m}$ from the chip surface, and then transported to the chip by an optical conveyor belt consisting of two counter-propagating Gaussian beams ($852\,\mathrm{nm}$)~\cite{Xu2023,Burgers2019,Xu2026}. Near the chip surface, both beams are focused to a waist of about $11\,\mathrm{\mu m}$ with powers of $128\,\mathrm{mW}$, forming a lattice of $1.25\,\mathrm{mK}$-depth standing-wave dipole traps. 

\begin{figure}[t]
\centering
\includegraphics[width=1.0\columnwidth]{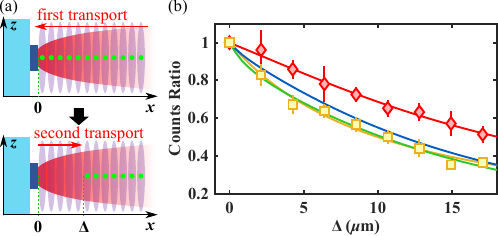}
\caption{(a) Illustration of the atom transportation process.  (b) Comparison of counts ratio $C(\Delta)$ between the cases of fluorescence collection light path modulated and unmodulated by the nanolensing effect, which are represented by yellow squares and red diamonds, respectively. The error bars represent the standard deviation of six experiments.  The red line is a fitting of red diamonds using expression \ref{CRw}, which gives the fitting parameters as $R=17.97\pm2.32\,\mathrm{\mu m}$ and $x_w=2.24\pm7.07\,\mathrm{\mu m}$. The yellow line is the same fitting of yellow squares, which gives the fitting parameters as $R=6.76\pm10.48\,\mathrm{\mu m}$ and $x_w=-7.32\pm9.82\,\mathrm{\mu m}$. The uncertainties are obtained with a $95\%$ confidence probability. The green line is a calculation result of expression \ref{CR1}, where the $|\mathbf{E}(x)|^2$ comes from a 2D numerical simulation of the nanolensing effect using a Gaussian beam with a Rayleigh length obtained by the blue line and is then multiplied by a correction factor $1/\sqrt{1+(\frac{x}{R})^2}$ to meet with the actual 3D situation. The blue line is a correction to the green line, considering the nanolensing effect on the conveyor beam.}
\label{fig3} 
\end{figure}

After transportation, a laser beam red-detuned by $2\mathrm{\pi} \times 28\,\mathrm{MHz}$ from the $\mathrm{\ket{F = 2}\,\rightarrow\ket{F' = 3}}$ transition of the $\mathrm{^{87}Rb\, D2}$ line is turned on to probe atoms. This probe beam has a linear polarization parallel to the z-axis and grazes on the chip with an incident angle of $85 ^\circ$, so about $80\%$ of the incident light power is reflected by the chip surface. We collect the reflected beam, turn its polarization by $90 ^\circ$ and retro-reflect it back to form a polarization-gradient cooling (PGC)~\cite{Chin2017}. Given that the chip has a width of $3.5\,\mathrm{mm}$, atoms within $300\,\mathrm{\mu m}$ of the chip surface can be uniformly excited. The fluorescence of the excited atoms is collected by the same objective used for focusing conveyor 1 beam and separated from the conveyor beam by a dichroic mirror (DM), and coupled into a single-mode fiber and detected by a single-photon counting module (SPCM)~\cite{SM}. Since the fluorescence collection light path passes across a waveguide, the collection should be modulated by the nanolens, as predicted in Fig.~\ref{fig1}(c).

\smallskip{}
\noindent \textbf{\emph{Results.-}}
To experimentally reveal the nanolensing effect, we exploit the precise positioning afforded by our optical conveyor belt. Figure~\ref{fig3}(a) illustrates the experimental procedure, where the purple lattices represent the dipole trap lattice generated by the conveyor belt, the green dots represent atoms, and the red pattern depicts the light field distribution of the fluorescence collection light path, modulated by the nanolensing effect. Cold atoms are first transported to the chip surface, where atoms contacting the surface immediately escape from the traps. Consequently, the atomic number density is a step function $\rho=\Theta(x)$, with $x$ the distance from the waveguide.
A second transport displaces the atom lattice by $\Delta$ in the $+x$ direction and purges atoms from the region $0<x<\Delta$~\cite{Xu2025,Kim2019}. Although we could not precisely manipulate single atoms to directly probe the nanolens, the step function produces a deterministic atom density function distribution that allows the investigation of the nanolens through statistical analysis of the atoms over many experimental trials. By recording the background-subtracted fluorescence counts during the first $10\, \mathrm{ms}$, we introduce the counts ratio 
\begin{equation}
    C(\Delta)=\frac{\int_{\Delta}^\infty \eta(x)\mathrm{d}x}{\int_{0}^\infty \eta(x)\mathrm{d}x}
\label{CR1}
\end{equation}
to quantify the modulation of collection for varied $\Delta$. Here, $\eta(x)$ is the fluorescence collection efficiency at $x$ that is proportional to the local field intensity ($\eta(x) \propto |\mathbf{E}(\mathbf{x})|^2$) according to the free-space beam quantum electrodynamics~\cite{Chen2024StandingWave,Chen2024Fluorescence}. Consequently, the derivative $-\mathrm{d}C/\mathrm{d}\Delta$ directly yields the normalized intensity profile of the collection mode and $C(\Delta)$ encodes its cumulative spatial distribution.

\begin{figure*}[t]
\centering
\includegraphics[width=1.0\textwidth]{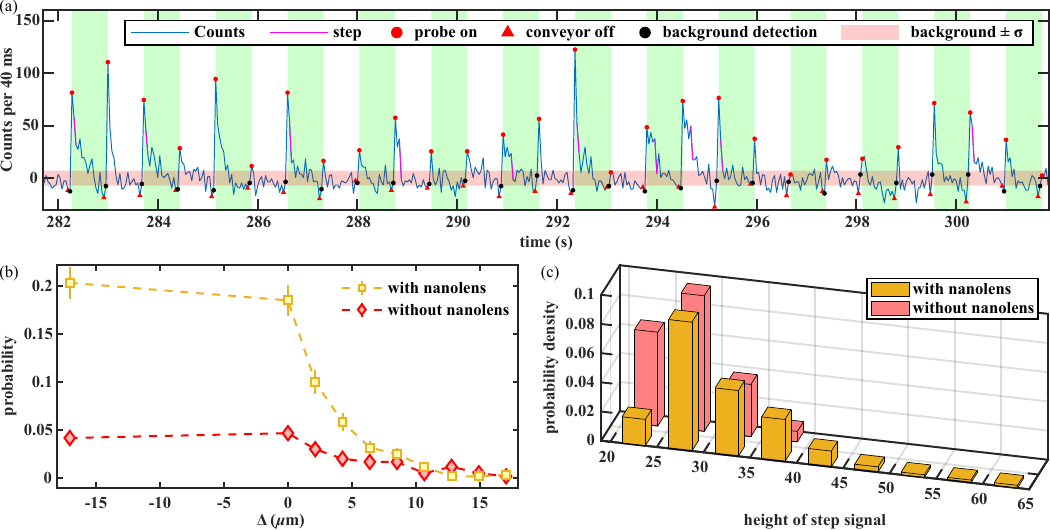}
\caption{(a) Typical step signal we get with nanolensing effect when $\Delta=0$. The blue line is part of the background-subtracted fluorescence counting track of our experiment. The fluorescence counts are collected in $40\,\mathrm{ms}$. The pink lines mark the step signal. The red circles and triangles mark the time of probe beam on and conveyor beams off, respectively. The black circles are the background detection time. The red shadow represents the region within one standard deviation from the mean value of background count. (b) The relation between the probability of single-atom signal and the second transport distance $\Delta$. The yellow squares and red diamonds are the probabilities of detecting a single-atom signal with and without the nanolensing effect, respectively. The error bars are given by the standard error of the binomial distribution with 600 measurements. (c) The comparison between the step height of the single-atom signal collected with and without the nanolensing effect. The yellow and red bars are the probability density distribution histograms of the single atom step height collected with and without the nanolensing effect, respectively.}
\label{fig4} 
\end{figure*}

Figure~\ref{fig3}(b) compares measured $C(\Delta)$ for the collection light path passing through the waveguide (yellow) and at a position $7.5\,\mathrm{\mu m}$ away from the waveguide (red), i.e., modulated and unmodulated by the nanolensing effect. It is found that $C(\Delta)$ obtained with the nanolensing effect falls faster as $\Delta$ increases, indicating a stronger focusing effect than an unmodulated Gaussian beam. By approximating the beam profile as a Gaussian beam, we have
\begin{equation}
    C(\Delta)=\frac{\mathrm{\pi}-2 \arctan(\frac{\Delta-x_w}{R})}{\mathrm{\pi}-2 \arctan(-\frac{x_w}{R})},
\label{CRw}
\end{equation}
where $R$ is the Rayleigh length and $x_w$ is the position of the waist. The direct fitting gives $R=6.76\pm10.48\,\mathrm{\mu m}$,  $x_w=-7.32\pm9.82\,\mathrm{\mu m}$ and $R=17.97\pm2.32\,\mathrm{\mu m}$, $x_w=2.24\pm7.07\,\mathrm{\mu m}$ for the cases with and without nanolens. The fitting for the nanolens case fails as the fitted $x_w$ significantly deviates from the numerical results of $250\,\mathrm{nm}$. Thus, we compute $C$ based on a 2D numerical simulation of the nanolensing effect, using an input Gaussian beam with a Rayleigh length obtained by the red line and then multiplying by a correction factor $1/\sqrt{1+({x}/{R})^2}$ to meet with the actual 3D situation. The blue line is a correction to the green line, considering the nanolensing effect on the conveyor belt beam~\cite{SM}, agreeing well with experimental data.

\smallskip{}
\noindent \textbf{\emph{Enhanced Single Atom Detection.-}}
The nanolensing effect directly translates into improved single-atom sensitivity. During probing, a single atom in the conveyor is heated and ejected within a few milliseconds, producing a characteristic step-like drop in the fluorescence trace~\cite{Frese2000,Hu2026}. Figure~\ref{fig4}(a) shows the typical step signal we get with the nanolensing effect when $\Delta=0$. We identify such single-atom events using threshold criteria on the step height, the post-step baseline, and the pre-step count level (see Supplemental Material for details~\cite{SM}). 

Figure~\ref{fig4}(b) shows the dependence of single-atom signal on $\Delta$. The yellow squares and red diamonds are the probabilities of detecting a single-atom signal with and without the nanolensing effect, respectively. With the nanolensing effect, the probability of detecting single atoms with $\Delta=0\,\mathrm{\mu m}$ is $20\%$, which is four times greater than without the nanolensing effect. Considering that the fluorescence counts collected with and without the nanolensing effect are almost the same~\cite{SM}, we can determine that the higher probability of single-atom signal is due to the higher fluorescence collection efficiency, not the greater number of atoms~\cite{Shadmany2025}. As $\Delta$ increases, the number of atoms in the region with high fluorescence collection efficiency is reduced, so the probability of detecting a single-atom signal decreases rapidly. From the yellow squares of $\Delta=0\,\mathrm{\mu m}$ and $\Delta=2.13\,\mathrm{\mu m}$, we can see that $50\%$ of the single-atom signals occur within about $2\,\mathrm{\mu m}$ of the waveguide surface, which is in agreement with the numerical simulation results that the waveguide lens mode is focused at a distance of $250\,\mathrm{nm}$ from the waveguide. As for the case of $\Delta<0$, the distribution of atom number density is the same as the case of $\Delta=0\,\mathrm{\mu m}$, so the probability of single-atom signal also remains at the same level. Figure~\ref{fig4}(c) compares the step height of the single-atom signal collected with and without the nanolensing effect. It is obvious that the nanolensing effect produces a long tail in the distribution of step height, confirming the great enhancement in the collection of single-atom fluorescence. This also indicates a tighter confinement of atoms in the dipole trap due to the nanolensing effect acting on the conveyor beam.

\smallskip{}
\noindent \textbf{\emph{Conclusion.-}}
We theoretically proposed and experimentally demonstrated the lensing effect induced by a single on-chip nanostructure, realized here with a dielectric waveguide. Using single atoms as sensitive probes, we directly mapped the subwavelength focal spot created by the waveguide nanostructure and observed a fourfold enhancement in the single atom detection efficiency compared to conventional Gaussian beam collection. Our approach establishes a novel paradigm for atom-nanophotonic interfaces, where nanophotonic elements serve the dual purpose of focusing light for atom trapping and enhancing light-matter coupling for efficient atom-photon interactions. Combining the waveguide nanolens with more complex photonic circuits, such as cavities, grating couplers, and beam splitters, could open up exciting opportunities for realizing advanced quantum devices and simulators with deterministic atom-photon interfaces. Our work lays the foundation for a novel class of hybrid nanophotonic-atomic systems that exploit nanoscale engineering to control and enhance the quantum dynamics of atoms and photons on a chip.

\smallskip{}
\begin{acknowledgments}
This work was funded by the National Key R\&D Program (Grant No. 2021YFA1402004), the National Natural Science Foundation of China (Grant Nos.~12134014, 92465201, 92265210, and 12293053). This work was also supported by the Fundamental Research Funds for the Central Universities, USTC Research Funds of the Double First-Class Initiative. The numerical calculations in this paper have been done on the supercomputing system in the Supercomputing Center of University of Science and Technology of China. This work was partially carried out at the USTC Center for Micro and Nanoscale Research and Fabrication.
\end{acknowledgments}

\bibliography{reference}

@misc{SM,
  howpublished={See the Supplemental Material for details.}
}

@article{Xu2025,
author = {Xu, Lei and Wang, Ling-Xiao and Chen, Guang-Jie and Wang, Zhu-Bo and Xu, Xin-Biao and Guo, Guang-Can and Zou, Chang-Ling and Xiang, Guo-Yong},
doi = {10.1103/rd53-4w5w},
issn = {2331-7019},
journal = {Physical Review Applied},
month = {aug},
number = {2},
pages = {024002},
publisher = {American Physical Society},
title = {{Dynamics of single atoms in optical tweezers near a chip's surface}},
url = {https://doi.org/10.1103/rd53-4w5w https://link.aps.org/doi/10.1103/rd53-4w5w},
volume = {24},
year = {2025}
}

@article{Hu2026,
archivePrefix = {arXiv},
arxivId = {2608.15637},
author = {Ya-Dong Hu and Tian-Yang Zhang and Dong-Qi Ma and Yi-Chen Zhang and Liang Chen and Wen-Yi Zhu and Hong-Jie Fan and Yan-Lei Zhang and Zhu-Bo Wang and Gang Li and Xi-Feng Ren and Guang-Can Guo and Chang-Ling Zou},
eprint = {2608.15637},
journal = {arXiv: 2608.15637},
month = {Aug},
title = {A scalable chip-integrated single-photon source array based on 50 individually addressable neutral atoms},
url = {https://arxiv.org/abs/2608.15637},
year = {2026}
}

@article{Wang2020,
  author    = {Wang, J. and Sciarrino, F. and Laing, A. and others},
  title     = {Integrated photonic quantum technologies},
  journal   = {Nature Photonics},
  volume    = {14},
  pages     = {273--284},
  year      = {2020},
  month     = {May},
  doi       = {10.1038/s41566-019-0532-1},
  url       = {https://doi.org/10.1038/s41566-019-0532-1},
  publisher = {Nature Publishing Group},
  received  = {10 May 2019},
  accepted  = {29 August 2019},
  published = {21 October 2019},
  issueDate = {May 2020}
}

@article{minzioni2019,
	title = {Roadmap on all-optical processing},
	volume = {21},
	issn = {2040-8978, 2040-8986},
	url = {https://iopscience.iop.org/article/10.1088/2040-8986/ab0e66},
	doi = {10.1088/2040-8986/ab0e66},
	number = {6},
	urldate = {2026-03-16},
	journal = {Journal of Optics},
	author = {Minzioni, Paolo and Lacava, Cosimo and Tanabe, Takasumi and Dong, Jianji and Hu, Xiaoyong and Csaba, Gyorgy and Porod, Wolfgang and Singh, Ghanshyam and Willner, Alan E and Almaiman, Ahmed and Torres-Company, Victor and Schröder, Jochen and Peacock, Anna C and Strain, Michael J and Parmigiani, Francesca and Contestabile, Giampiero and Marpaung, David and Liu, Zhixin and Bowers, John E and Chang, Lin and Fabbri, Simon and Ramos Vázquez, María and Bharadwaj, Vibhav and Eaton, Shane M and Lodahl, Peter and Zhang, Xiang and Eggleton, Benjamin J and Munro, William John and Nemoto, Kae and Morin, Olivier and Laurat, Julien and Nunn, Joshua},
	month = {jun},
	year = {2019},
	pages = {063001},
}

@article{Chang2014,
  author    = {Chang, D. and Vuleti\'{c}, V. and Lukin, M.},
  title     = {Quantum nonlinear optics — photon by photon},
  journal   = {Nature Photonics},
  volume    = {8},
  pages     = {685--694},
  year      = {2014},
  month     = {September},
  doi       = {10.1038/nphoton.2014.192},
  url       = {https://doi.org/10.1038/nphoton.2014.192},
  publisher = {Nature Publishing Group},
  received  = {20 February 2014},
  accepted  = {22 July 2014},
  published = {24 August 2014},
  issueDate = {September 2014}
}

@article{Liu2023,
  title = {Realization of Strong Coupling between Deterministic Single-Atom Arrays and a High-Finesse Miniature Optical Cavity},
  author = {Liu, Yanxin and Wang, Zhihui and Yang, Pengfei and Wang, Qinxia and Fan, Qing and Guan, Shijun and Li, Gang and Zhang, Pengfei and Zhang, Tiancai},
  journal = {Phys. Rev. Lett.},
  volume = {130},
  issue = {17},
  pages = {173601},
  numpages = {7},
  year = {2023},
  month = {Apr},
  publisher = {American Physical Society},
  doi = {10.1103/PhysRevLett.130.173601},
  url = {https://link.aps.org/doi/10.1103/PhysRevLett.130.173601}
}

@article{Yang2023,
author = {Yang, Pengfei and Li, Ming and Han, Xing and He, Hai and Li, Gang and Zou, Chang-Ling and Zhang, Pengfei and Qian, Yuhua and Zhang, Tiancai},
title = {Non-Reciprocal Cavity Polariton with Atoms Strongly Coupled to Optical Cavity},
journal = {Laser \& Photonics Reviews},
volume = {17},
number = {7},
pages = {2200574},
doi = {https://doi.org/10.1002/lpor.202200574},
url = {https://onlinelibrary.wiley.com/doi/abs/10.1002/lpor.202200574},
year = {2023}
}

@article{Specht2011,
  author    = {Specht, H. and N{\"o}lleke, C. and Reiserer, A. and others},
  title     = {A single-atom quantum memory},
  journal   = {Nature},
  volume    = {473},
  pages     = {190--193},
  year      = {2011},
  month     = {May},
  doi       = {10.1038/nature09997},
  url       = {https://doi.org/10.1038/nature09997},
  issue     = {7346},
  publisher = {Nature Publishing Group},
  received  = {20 January 2011},
  accepted  = {09 March 2011},
  published = {01 May 2011},
  issueDate = {12 May 2011}
}

@article{Corzo2019,
	title = {Waveguide-coupled single collective excitation of atomic arrays},
	volume = {566},
	issn = {0028-0836, 1476-4687},
	url = {https://www.nature.com/articles/s41586-019-0902-3},
	doi = {10.1038/s41586-019-0902-3},
	number = {7744},
	urldate = {2026-03-16},
	journal = {Nature},
	author = {Corzo, Neil V. and Raskop, Jérémy and Chandra, Aveek and Sheremet, Alexandra S. and Gouraud, Baptiste and Laurat, Julien},
	month = {feb},
	year = {2019},
	pages = {359--362},
}

@article{Javadi2015,
  author    = {Javadi, A. and S{\"o}llner, I. and Arcari, M. and others},
  title     = {Single-photon non-linear optics with a quantum dot in a waveguide},
  journal   = {Nature Communications},
  volume    = {6},
  pages     = {8655},
  year      = {2015},
  doi       = {10.1038/ncomms9655},
  url       = {https://doi.org/10.1038/ncomms9655},
  publisher = {Nature Publishing Group},
  received  = {11 April 2015},
  accepted  = {17 September 2015},
  published = {23 October 2015}
}

@article{Hacker2016,
  author    = {Hacker, B. and Welte, S. and Rempe, G. and others},
  title     = {A photon–photon quantum gate based on a single atom in an optical resonator},
  journal   = {Nature},
  volume    = {536},
  pages     = {193--196},
  year      = {2016},
  month     = {August},
  doi       = {10.1038/nature18592},
  url       = {https://doi.org/10.1038/nature18592},
  publisher = {Nature Publishing Group},
  received  = {19 February 2016},
  accepted  = {12 May 2016},
  published = {06 July 2016},
  issueDate = {11 August 2016}
}

@article{Grinkemeyer2025,
author = {Brandon Grinkemeyer and Elmer Guardado-Sanchez and Ivana Dimitrova and Danilo Shchepanovich and G. Eirini Mandopoulou and Johannes Borregaard and Vladan Vuletić and Mikhail D. Lukin },
title = {Error-detected quantum operations with neutral atoms mediated by an optical cavity},
journal = {Science},
volume = {387},
number = {6740},
pages = {1301-1305},
year = {2025},
doi = {10.1126/science.adr7075},
URL = {https://www.science.org/doi/abs/10.1126/science.adr7075}
}

@article{Bhaskar2017,
  author    = {Bhaskar, M. K. and Sukachev, D. D. and Sipahigil, A. and others},
  title     = {Quantum nonlinear optics with a germanium-vacancy color center in a nanoscale diamond waveguide},
  journal   = {Physical Review Letters},
  volume    = {118},
  number    = {22},
  pages     = {223603},
  year      = {2017},
  doi       = {10.1103/PhysRevLett.118.223603},
  publisher = {American Physical Society}
}

@article{Kalb2015,
  title = {Heralded Storage of a Photonic Quantum Bit in a Single Atom},
  author = {Kalb, Norbert and Reiserer, Andreas and Ritter, Stephan and Rempe, Gerhard},
  journal = {Phys. Rev. Lett.},
  volume = {114},
  issue = {22},
  pages = {220501},
  numpages = {5},
  year = {2015},
  month = {Jun},
  publisher = {American Physical Society},
  doi = {10.1103/PhysRevLett.114.220501},
  url = {https://link.aps.org/doi/10.1103/PhysRevLett.114.220501}
}

@article{Waldermann2007,
  author    = {F. C. Waldermann and P. Olivero and J. Nunn and K. Surmacz and Z. Y. Wang and D. Jaksch and R. A. Taylor and I. A. Walmsley and M. Draganski and P. Reichart and A. D. Greentree and D. N. Jamieson and S. Prawer},
  title     = {Creating diamond color centers for quantum optical applications},
  journal   = {Diamond and Related Materials},
  volume    = {16},
  number    = {11},
  pages     = {1887--1895},
  year      = {2007},
  issn      = {0925-9635},
  doi       = {10.1016/j.diamond.2007.09.009},
  publisher = {Elsevier}
}

@article{Andreas2022,
  title = {Colloquium: Cavity-enhanced quantum network nodes},
  author = {Reiserer, Andreas},
  journal = {Rev. Mod. Phys.},
  volume = {94},
  issue = {4},
  pages = {041003},
  numpages = {22},
  year = {2022},
  month = {Dec},
  publisher = {American Physical Society},
  doi = {10.1103/RevModPhys.94.041003},
  url = {https://link.aps.org/doi/10.1103/RevModPhys.94.041003}
}

@article{Hartung2024,
author = {Lukas Hartung  and Matthias Seubert  and Stephan Welte  and Emanuele Distante  and Gerhard Rempe },
title = {A quantum-network register assembled with optical tweezers in an optical cavity},
journal = {Science},
volume = {385},
number = {6705},
pages = {179-183},
year = {2024},
doi = {10.1126/science.ado6471},
URL = {https://www.science.org/doi/abs/10.1126/science.ado6471},
}

@article{hu2025,
	title = {Site-{Selective} {Cavity} {Readout} and {Classical} {Error} {Correction} of a 5-{Bit} {Atomic} {Register}},
	volume = {134},
	issn = {0031-9007, 1079-7114},
	url = {https://link.aps.org/doi/10.1103/PhysRevLett.134.120801},
	doi = {10.1103/PhysRevLett.134.120801},
	number = {12},
	urldate = {2026-03-04},
	journal = {Physical Review Letters},
	author = {Hu, Beili and Sinclair, Josiah and Bytyqi, Edita and Chong, Michelle and Rudelis, Alyssa and Ramette, Joshua and Vendeiro, Zachary and Vuletić, Vladan},
	month = {mar},
	year = {2025},
	pages = {120801},
}

@article{Pelucchi2022,
  author    = {Pelucchi, E. and Fagas, G. and Aharonovich, I. and others},
  title     = {The potential and global outlook of integrated photonics for quantum technologies},
  journal   = {Nature Reviews Physics},
  volume    = {4},
  pages     = {194--208},
  year      = {2022},
  month     = {March},
  doi       = {10.1038/s42254-021-00398-z},
  url       = {https://doi.org/10.1038/s42254-021-00398-z},
  publisher = {Nature Publishing Group},
  accepted  = {04 November 2021},
  published = {23 December 2021},
  issueDate = {March 2022}
}

@article{Wang2025,
	title = {Purcell-{Enhanced} {Generation} of {Photonic} {Bell} {States} via the {Inelastic} {Scattering} off {Single} {Atoms}},
	volume = {134},
	issn = {0031-9007, 1079-7114},
	url = {https://link.aps.org/doi/10.1103/PhysRevLett.134.053401},
	doi = {10.1103/PhysRevLett.134.053401},
	number = {5},
	urldate = {2026-03-04},
	journal = {Physical Review Letters},
	author = {Wang, Jian and Zhou, Xiao-Long and Shen, Ze-Min and Huang, Dong-Yu and He, Si-Jian and Huang, Qi-Yang and Liu, Yi-Jia and Li, Chuan-Feng and Guo, Guang-Can},
	month = {feb},
	year = {2025},
	pages = {053401},
}

@article{chang2019,
	title = {Microring resonators on a suspended membrane circuit for atom–light interactions},
	volume = {6},
	issn = {2334-2536},
	url = {https://opg.optica.org/abstract.cfm?URI=optica-6-9-1203},
	doi = {10.1364/OPTICA.6.001203},
	number = {9},
	urldate = {2025-11-12},
	journal = {Optica},
	author = {Chang, Tzu-Han and Fields, Brian M. and Kim, May E. and Hung, Chen-Lung},
	month = {sep},
	year = {2019},
	pages = {1203}
}

@article{Zhou2023,
  title = {Coupling Single Atoms to a Nanophotonic Whispering-Gallery-Mode Resonator via Optical Guiding},
  author = {Zhou, Xinchao and Tamura, Hikaru and Chang, Tzu-Han and Hung, Chen-Lung},
  journal = {Phys. Rev. Lett.},
  volume = {130},
  issue = {10},
  pages = {103601},
  numpages = {7},
  year = {2023},
  month = {Mar},
  publisher = {American Physical Society},
  doi = {10.1103/PhysRevLett.130.103601},
  url = {https://link.aps.org/doi/10.1103/PhysRevLett.130.103601}
}

@article{Menon2024,
  author    = {Menon, S. G. and Glachman, N. and Pompili, M. and others},
  title     = {An integrated atom array-nanophotonic chip platform with background-free imaging},
  journal   = {Nature Communications},
  volume    = {15},
  pages     = {6156},
  year      = {2024},
  doi       = {10.1038/s41467-024-50355-4},
  url       = {https://doi.org/10.1038/s41467-024-50355-4},
  publisher = {Nature Publishing Group},
  received  = {23 February 2024},
  accepted  = {09 July 2024},
  published = {22 July 2024}
}

@article{Kim2020,
  author    = {Je-Hyung Kim and Shahriar Aghaeimeibodi and Jacques Carolan and Dirk Englund and Edo Waks},
  title     = {Hybrid integration methods for on-chip quantum photonics},
  journal   = {Optica},
  volume    = {7},
  pages     = {291--308},
  year      = {2020},
  doi       = {10.1364/OPTICA.384118},
  publisher = {Optica Publishing Group}
}

@article{Wan2020,
  author    = {Wan, N. H. and Lu, T. J. and Chen, K. C. and others},
  title     = {Large-scale integration of artificial atoms in hybrid photonic circuits},
  journal   = {Nature},
  volume    = {583},
  pages     = {226--231},
  year      = {2020},
  month     = {July},
  doi       = {10.1038/s41586-020-2441-3},
  url       = {https://doi.org/10.1038/s41586-020-2441-3},
  publisher = {Nature Publishing Group},
  received  = {13 October 2019},
  accepted  = {02 April 2020},
  published = {08 July 2020},
  issueDate = {09 July 2020}
}

@article{Kaufman2021,
  author    = {Kaufman, A. M. and Ni, K. K.},
  title     = {Quantum science with optical tweezer arrays of ultracold atoms and molecules},
  journal   = {Nature Physics},
  volume    = {17},
  pages     = {1324--1333},
  year      = {2021},
  month     = {December},
  doi       = {10.1038/s41567-021-01357-2},
  url       = {https://doi.org/10.1038/s41567-021-01357-2},
  publisher = {Nature Publishing Group},
  received  = {05 September 2020},
  accepted  = {04 August 2021},
  published = {11 November 2021},
  issueDate = {December 2021}
}

@article{Dordevic2021,
  author    = {\DJ or\dj evi\'{c}, T. and Samutpraphoot, P. and Ocola, P. L. and others},
  title     = {Entanglement transport and a nanophotonic interface for atoms in optical tweezers},
  journal   = {Science},
  volume    = {373},
  number    = {6562},
  pages     = {1511--1514},
  year      = {2021},
  doi       = {10.1126/science.abi9917},
  publisher = {American Association for the Advancement of Science}
}

@article{Thompson2013,
  author    = {J. D. Thompson and T. G. Tiecke and N. P. de Leon and others},
  title     = {Coupling a Single Trapped Atom to a Nanoscale Optical Cavity},
  journal   = {Science},
  volume    = {340},
  number    = {6137},
  pages     = {1202--1205},
  year      = {2013},
  doi       = {10.1126/science.1237125},
  publisher = {American Association for the Advancement of Science}
}

@article{Seubert2025,
  title = {Tweezer-Assisted Subwavelength Positioning of Atomic Arrays in an Optical Cavity},
  author = {Seubert, M. and Hartung, L. and Welte, S. and Rempe, G. and Distante, E.},
  journal = {PRX Quantum},
  volume = {6},
  issue = {1},
  pages = {010322},
  numpages = {10},
  year = {2025},
  month = {Feb},
  publisher = {American Physical Society},
  doi = {10.1103/PRXQuantum.6.010322},
  url = {https://link.aps.org/doi/10.1103/PRXQuantum.6.010322}
}

@article{Brown2023,
	title = {Time-of-flight quantum tomography of an atom in an optical tweezer},
	volume = {19},
	issn = {1745-2473, 1745-2481},
	url = {https://www.nature.com/articles/s41567-022-01890-8},
	doi = {10.1038/s41567-022-01890-8},
	number = {4},
	urldate = {2025-11-12},
	journal = {Nature Physics},
	author = {Brown, M. O. and Muleady, S. R. and Dworschack, W. J. and Lewis-Swan, R. J. and Rey, A. M. and Romero-Isart, O. and Regal, C. A.},
	month = {apr},
	year = {2023},
	pages = {569--573},
}

@article{Zhou2024,
  title = {Trapped Atoms and Superradiance on an Integrated Nanophotonic Microring Circuit},
  author = {Zhou, Xinchao and Tamura, Hikaru and Chang, Tzu-Han and Hung, Chen-Lung},
  journal = {Phys. Rev. X},
  volume = {14},
  issue = {3},
  pages = {031004},
  numpages = {11},
  year = {2024},
  month = {Jul},
  publisher = {American Physical Society},
  doi = {10.1103/PhysRevX.14.031004},
  url = {https://link.aps.org/doi/10.1103/PhysRevX.14.031004}
}

@article{Luan2020,
author = {Luan, Xingsheng and B\'{e}guin, Jean-Baptiste and Burgers, Alex P. and Qin, Zhongzhong and Yu, Su-Peng and Kimble, Harry J.},
title = {The Integration of Photonic Crystal Waveguides with Atom Arrays in Optical Tweezers},
journal = {Advanced Quantum Technologies},
volume = {3},
number = {11},
pages = {2000008},
doi = {https://doi.org/10.1002/qute.202000008},
url = {https://onlinelibrary.wiley.com/doi/abs/10.1002/qute.202000008},
year = {2020}
}

@article{Kim2019,
  author    = {Kim, M. E. and Chang, T. H. and Fields, B. M. and others},
  title     = {Trapping single atoms on a nanophotonic circuit with configurable tweezer lattices},
  journal   = {Nature Communications},
  volume    = {10},
  pages     = {1647},
  year      = {2019},
  doi       = {10.1038/s41467-019-09635-7},
  url       = {https://doi.org/10.1038/s41467-019-09635-7},
  publisher = {Nature Publishing Group},
  received  = {30 October 2018},
  accepted  = {20 March 2019},
  published = {09 April 2019}
}

@article{Meng2020,
  title = {Imaging and Localizing Individual Atoms Interfaced with a Nanophotonic Waveguide},
  author = {Meng, Y. and Liedl, C. and Pucher, S. and Rauschenbeutel, A. and Schneeweiss, P.},
  journal = {Phys. Rev. Lett.},
  volume = {125},
  issue = {5},
  pages = {053603},
  numpages = {6},
  year = {2020},
  month = {Jul},
  publisher = {American Physical Society},
  doi = {10.1103/PhysRevLett.125.053603},
  url = {https://link.aps.org/doi/10.1103/PhysRevLett.125.053603}
}

@article{Will2021,
	title = {Coupling a {Single} {Trapped} {Atom} to a {Whispering}-{Gallery}-{Mode} {Microresonator}},
	volume = {126},
	issn = {0031-9007, 1079-7114},
	url = {https://link.aps.org/doi/10.1103/PhysRevLett.126.233602},
	doi = {10.1103/PhysRevLett.126.233602},
	number = {23},
	urldate = {2025-11-12},
	journal = {Physical Review Letters},
	author = {Will, Elisa and Masters, Luke and Rauschenbeutel, Arno and Scheucher, Michael and Volz, Jürgen},
	month = {jun},
	year = {2021},
	pages = {233602},
}

@article{Samutpraphoot2020,
	title = {Strong {Coupling} of {Two} {Individually} {Controlled} {Atoms} via a {Nanophotonic} {Cavity}},
	volume = {124},
	issn = {0031-9007, 1079-7114},
	url = {https://link.aps.org/doi/10.1103/PhysRevLett.124.063602},
	doi = {10.1103/PhysRevLett.124.063602},
	number = {6},
	urldate = {2025-11-12},
	journal = {Physical Review Letters},
	author = {Samutpraphoot, Polnop and Đorđević, Tamara and Ocola, Paloma L. and Bernien, Hannes and Senko, Crystal and Vuletić, Vladan and Lukin, Mikhail D.},
	month = {feb},
	year = {2020},
	pages = {063602},
}

@article{Han2026,
	title = {Interfacing of an optical nanofiber with tunably spaced atoms in an optical lattice},
	url = {http://arxiv.org/abs/2509.22958},
	doi = {10.48550/arXiv.2509.22958},
	urldate = {2026-03-04},
	publisher = {arXiv},
	author = {Han, Hyok Sang and Lee, Ahreum and Subhankar, Sarthak and Fatemi, Fredrik K. and Rolston, S. L.},
	month = {jan},
	year = {2026},
    journal = {arXiv:2509.22958},
}

@article{Liu2022,
  title = {Proposal for low-power atom trapping on a GaN-on-sapphire chip},
  author = {Liu, Aiping and Xu, Lei and Xu, Xin-Biao and Chen, Guang-Jie and Zhang, Pengfei and Xiang, Guo-Yong and Guo, Guang-Can and Wang, Qin and Zou, Chang-Ling},
  journal = {Phys. Rev. A},
  volume = {106},
  issue = {3},
  pages = {033104},
  numpages = {9},
  year = {2022},
  month = {Sep},
  publisher = {American Physical Society},
  doi = {10.1103/PhysRevA.106.033104},
  url = {https://link.aps.org/doi/10.1103/PhysRevA.106.033104}
}

@article{Zheng2022,
author = {Zheng, Yanzhen and Sun, Changzheng and Xiong, Bing and Wang, Lai and Hao, Zhibiao and Wang, Jian and Han, Yanjun and Li, Hongtao and Yu, Jiadong and Luo, Yi},
title = {Integrated Gallium Nitride Nonlinear Photonics},
journal = {Laser \& Photonics Reviews},
volume = {16},
number = {1},
pages = {2100071},
doi = {https://doi.org/10.1002/lpor.202100071},
url = {https://onlinelibrary.wiley.com/doi/abs/10.1002/lpor.202100071},
year = {2022}
}

@article{Xu2023,
  author    = {Lei Xu and Ling-Xiao Wang and Guang-Jie Chen and Liang Chen and Yuan-Hao Yang and Xin-Biao Xu and Aiping Liu and Chuan-Feng Li and Guang-Can Guo and Chang-Ling Zou and Guo-Yong Xiang},
  title     = {Transporting Cold Atoms towards a GaN-on-Sapphire Chip via an Optical Conveyor Belt},
  journal   = {Chinese Physics Letters},
  volume    = {40},
  number    = {9},
  pages     = {093701},
  year      = {2023},
  doi       = {10.1088/0256-307X/40/9/093701},
  publisher = {IOP Publishing}
}

@article{Burgers2019,
author = {A. P. Burgers  and L. S. Peng  and J. A. Muniz  and A. C. McClung  and M. J. Martin  and H. J. Kimble },
title = {Clocked atom delivery to a photonic crystal waveguide},
journal = {Proceedings of the National Academy of Sciences},
volume = {116},
number = {2},
pages = {456-465},
year = {2019},
doi = {10.1073/pnas.1817249115},
URL = {https://www.pnas.org/doi/abs/10.1073/pnas.1817249115},
}

@article{Chin2017,
  title = {Polarization gradient cooling of single atoms in optical dipole traps},
  author = {Chin, Yue-Sum and Steiner, Matthias and Kurtsiefer, Christian},
  journal = {Phys. Rev. A},
  volume = {96},
  issue = {3},
  pages = {033406},
  numpages = {5},
  year = {2017},
  month = {Sep},
  publisher = {American Physical Society},
  doi = {10.1103/PhysRevA.96.033406},
  url = {https://link.aps.org/doi/10.1103/PhysRevA.96.033406}
}

@article{Chen2024StandingWave,
author = {Guang-Jie Chen and Jun-Jie Wang and Zhu-Bo Wang and Dong Zhao and Yan-Lei Zhang and Ai-Ping Liu and Chun-Hua Dong and Kun Huang and Guang-Can Guo and Chang-Ling Zou},
journal = {Optics Express},
number = {22},
pages = {39039--39052},
publisher = {Optica Publishing Group},
title = {Standing-wave atom tweezer},
volume = {32},
month = {Oct},
year = {2024},
url = {https://opg.optica.org/oe/abstract.cfm?URI=oe-32-22-39039},
doi = {10.1364/OE.538445},
}

@article{Chen2024Fluorescence,
author = {Guang-Jie Chen and Jun-Jie Wang and Ya-Nan Lv and Hong-Jie Fan and Zhu-Bo Wang and Gang Li and Chun-Hua Dong and Yan-Lei Zhang and Guang-Can Guo and Chang-Ling Zou},
journal = {Optics Letters},
number = {17},
pages = {5011--5014},
publisher = {Optica Publishing Group},
title = {Fluorescence collection efficiency of atoms in dipole traps},
volume = {49},
month = {Sep},
year = {2024},
url = {https://opg.optica.org/ol/abstract.cfm?URI=ol-49-17-5011},
doi = {10.1364/OL.537054},
}

@article{Frese2000,
	title = {Single {Atoms} in an {Optical} {Dipole} {Trap}: {Towards} a {Deterministic} {Source} of {Cold} {Atoms}},
	volume = {85},
	copyright = {http://link.aps.org/licenses/aps-default-license},
	issn = {0031-9007, 1079-7114},
	shorttitle = {Single {Atoms} in an {Optical} {Dipole} {Trap}},
	url = {https://link.aps.org/doi/10.1103/PhysRevLett.85.3777},
	doi = {10.1103/PhysRevLett.85.3777},
	number = {18},
	urldate = {2025-11-12},
	journal = {Physical Review Letters},
	author = {Frese, D. and Ueberholz, B. and Kuhr, S. and Alt, W. and Schrader, D. and Gomer, V. and Meschede, D.},
	month = {oct},
	year = {2000},
	pages = {3777--3780},
}

@article{Shadmany2025,
author = {Danial Shadmany  and Aishwarya Kumar  and Anna Soper  and Lukas Palm  and Chuan Yin  and Henry Ando  and Bowen Li  and Lavanya Taneja  and Matt Jaffe  and Schuster David  and Jon Simon },
title = {Cavity {QED} in a high {NA} resonator},
journal = {Science Advances},
volume = {11},
number = {9},
pages = {eads8171},
year = {2025},
doi = {10.1126/sciadv.ads8171},
URL = {https://www.science.org/doi/abs/10.1126/sciadv.ads8171},
}

@article{Xu2026,
  title = {Efficient single-atom transfer from an optical conveyor belt to a tightly confined optical tweezer},
  author = {Xu, Lei and Wang, Ling-Xiao and Chen, Guang-Jie and Wang, Zhu-Bo and Xu, Xin-Biao and Guo, Guang-Can and Zou, Chang-Ling and Xiang, Guo-Yong},
  journal = {Phys. Rev. Appl.},
  volume = {26},
  issue = {2},
  pages = {024037},
  numpages = {8},
  year = {2026},
  month = {Aug},
  publisher = {American Physical Society},
  doi = {10.1103/wccv-wpqh},
  url = {https://link.aps.org/doi/10.1103/wccv-wpqh}
}

\end{document}


\title{Supplemental Material for:\\Lensing and enhanced single atom detection via a single on-chip nanostructure}

\begin{center}
{\Large \textbf{Supplemental Material for:\\
Lensing and enhanced single atom detection via a single-pixel nanostructure}}
\end{center}
\vspace{1em}

\section{Appendix A: Numerical simulation of nanolensing effect}
\begin{figure}[hbt]
\centering
\includegraphics[width=0.95\textwidth]{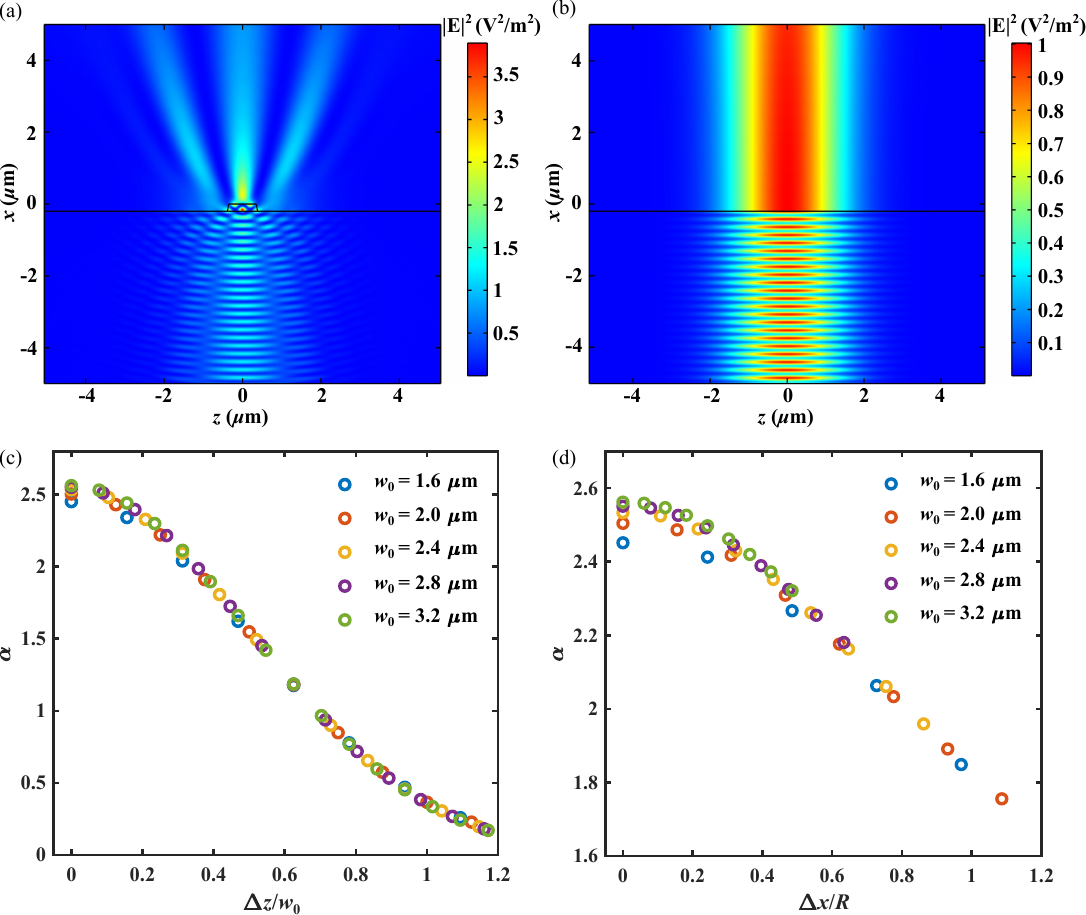}
\caption{The numerical simulation result of the nanolensing effect. (a) The simulated $|\mathbf{E}|^2$ distribution of the total field in the vicinity of a trapezium dielectric waveguide (width $a=700\, \mathrm{nm}$, thickness $t = 200\,\mathrm{nm}$, refractive index $n = 2.3566$) illuminated from below by a Gaussian beam with a wavelength of $\lambda = 780\,\mathrm{nm}$ and a waist of $2\,\mathrm{\mu m}$.
Below the horizontal straight line in the picture is the sapphire substrate, and above it is the air.
(b) The result of the same Gaussian beam not passing through the waveguide. (c)(d) The relation between $\alpha$ and the position deviation of the incident Gaussian beam. $\Delta z$ in (c) and $\Delta x$ in (d) are the position deviations of the incident Gaussian beam in radial and axial directions, respectively. Different colors represent different incident beam waists ($w_0$). $R$ is the Rayleigh length of the incident beam.}
\label{figS1}
\end{figure}

Figure~\ref{figS1} shows the numerical simulation result of the nanolensing effect.
Figure~\ref{figS1}(a) shows the simulated $|\mathbf{E}|^2$ distribution of the total field in the vicinity of a trapezium dielectric waveguide (width $a=700\, \mathrm{nm}$, thickness $t = 200\,\mathrm{nm}$, refractive index $n = 2.3566$) illuminated from below by a Gaussian beam with a wavelength of $\lambda = 780\,\mathrm{nm}$ and a waist of $2\,\mathrm{\mu m}$.
Below the horizontal straight line in the picture is the sapphire substrate, and above it is the air.
The result of the same Gaussian beam not passing through the waveguide is shown in Fig.~\ref{figS1}(b).

Then we investigate the relation between $\alpha$ and the position deviation of the incident Gaussian beam. 
$\Delta z$ in (c) and $\Delta x$ in (d) are the position deviations of the incident Gaussian beam in radial and axial directions, respectively.
Different colors represent different incident beam waists, ($w_0$).
$R$ is the Rayleigh length of the incident beam.
When $\Delta z=\Delta x=0$, $\alpha$ has a maximum of 2.5 approximately, and it hardly changes with the incident beam waist.
In the radial direction, $\alpha$ has a half-width at half-maximum (HWHM) of $w_0/2$.
In the axial direction, the HWHM of $\alpha$ is much greater than $R$.
For a relatively large incident beam waist, there can be a large alignment tolerance, and the focusing effect is unchanged.

\section{Appendix B: Time sequence of the experiment}

\begin{figure}[h]
\centering
\includegraphics[width=1.0\textwidth]{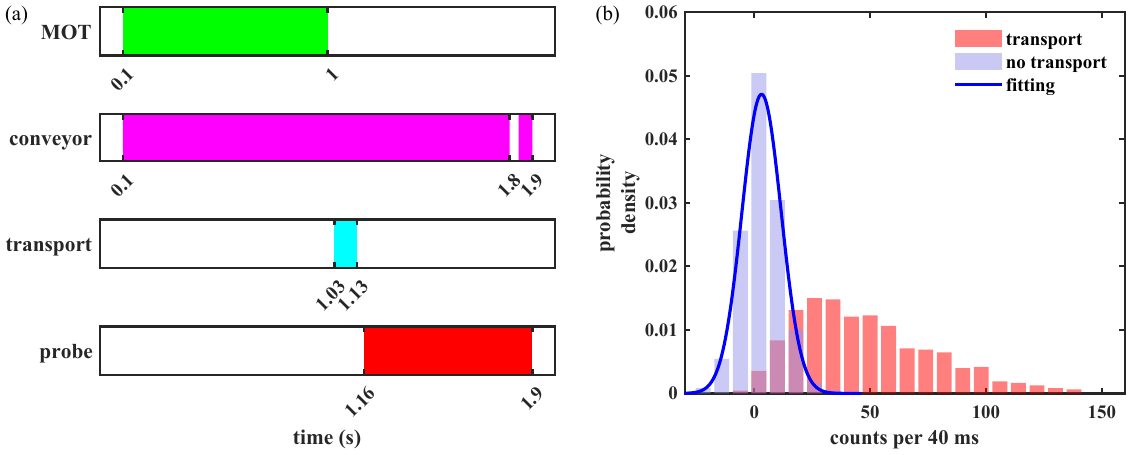}
\caption{(a) Time sequence of the experiment. (b) Comparison of probability density distributions of background noise-subtracted fluorescence counts whether atoms are transported to the chip surface. Blue line: a normal distribution fitting of the blue histogram. The fitting gives a mean value of $\mu_{bg}=3.11$ and a standard deviation of $\sigma_{bg}=8.48$, which means no atom fluorescence is collected.}
\label{figS2}
\end{figure}

Figure~\ref{figS2}(a) shows the time sequence of our experiment.
The probe process begins at $1.16\,\mathrm{s}$.
The conveyor beams are shut down at $1.8\,\mathrm{s}$ to release all the atoms and then re-opened at $1.84\,\mathrm{s}$ to obtain the background noise.
Fig.~\ref{figS2}(b) shows the comparison of probability density distributions of background-noise-subtracted fluorescence counts whether atoms are transported to the chip surface.
The red histogram is the probability density distribution of the collected photon counts in the first $40 \mathrm{ms}$ of probe process after the cold atoms in the MOT are transported to the area near the chip surface.
The blue histogram is the probability density distribution of the first $40\,\mathrm{ms}$ photon counts that atoms are not transported, for comparison.
The background has already been subtracted in both cases.
The blue line is a normal distribution fitting of the blue histogram. The fitting gives a mean value of $\mu_{bg}=3.11$ and a standard deviation of $\sigma_{bg}=8.48$, which means no atom fluorescence is collected.
The fluorescence signal is significantly larger in the case that atoms are transported to the chip surface, which proves the ability to transport atoms and detect atom fluorescence in our system.

\section{Appendix C: Alignment of the fluorescence collection light path}

\begin{figure}[hbt]
\centering
\includegraphics[width=0.9\columnwidth]{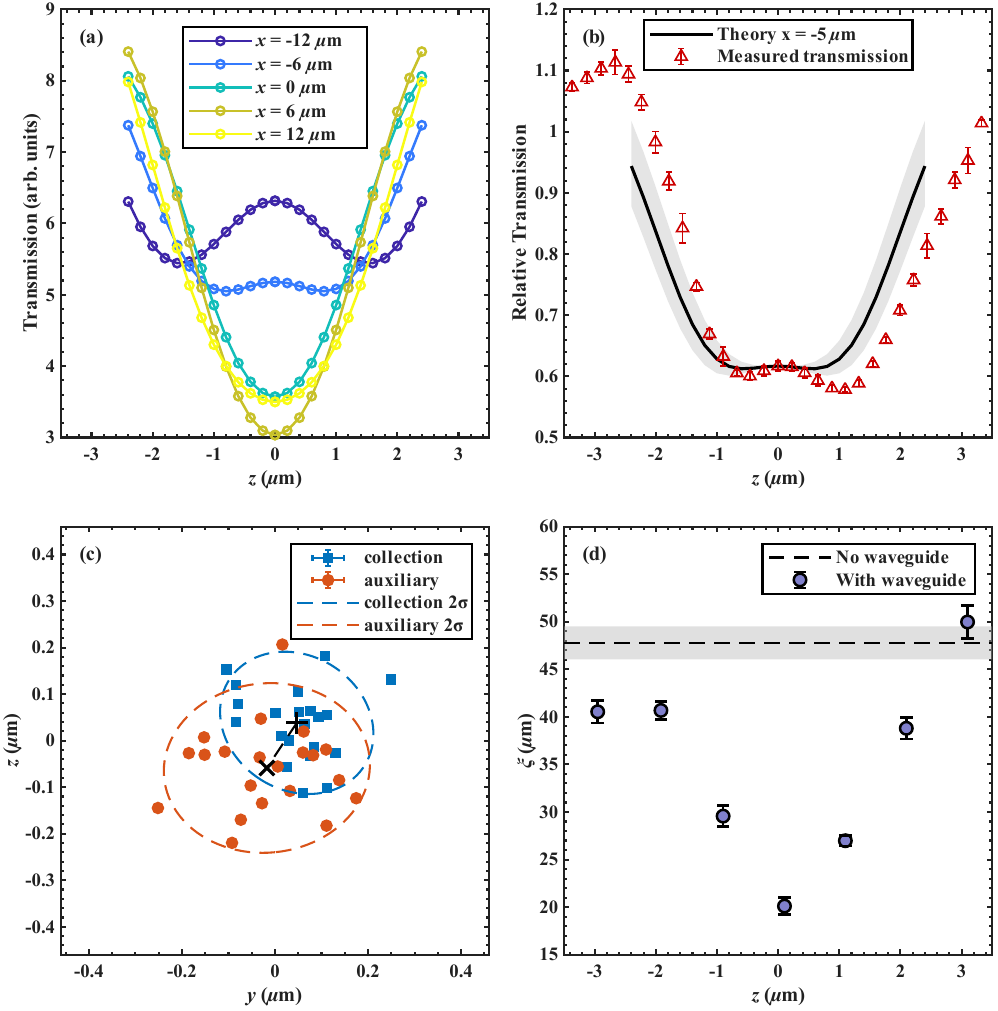}
\caption{
(a) Simulated transmission versus the auxiliary beam waist position $(x,z)$, calculated using the angular spectrum method. (b) The red triangles are experimentally measured relative transmission versus $z$. The black solid line is the simulated transmission for the waveguide distance $x = -5\,\mathrm{\mu m}$, which best matches the experimental data, selected by least-squares fitting. The gray shaded region shows the envelope bounded by the simulations for the $x = -5\pm1\,\mathrm{\mu m}$. (c) Centers of the back-injected collection spot (blue squares) and the 852 nm auxiliary focusing beam spot (red circles) with $2\sigma$ confidence ellipses, offset by $(x_0,y_0)=(452,247.8)\;\mu$m. The two beams share the same objective. The distance between their centers is less than $0.3\;\mu$m at a 95.6\% confidence level. (d) Decay length $\xi$ of the counts ratio versus $z$, extracted by fitting $\exp(-\Delta/\xi)$ to the counts ratio data. The gray band shows $\xi$ without the waveguide ($\pm1\sigma$), and the blue circles show $\xi$ with the waveguide. The nanolensing effect is present only when the alignment error is less than $2\;\mu$m.
}
\label{fig:S3}
\end{figure}

The nanolensing effect relies on the spatial alignment between the fluorescence collection light path and the waveguide. In this appendix, we present experimental characterizations that confirm this alignment.

First, we align an auxiliary $852\,\mathrm{nm}$ beam and the fluorescence collection path so that their optical axes and beam waists coincide. Then we adjust the common mirror before the objective and collect the transmitted light of the auxiliary beam passing across the waveguide. To avoid power saturation, a 1 mm diameter aperture is placed before the photoelectric detector. By monitoring the transmission of the auxiliary beam, we determine the relative position between the fluorescence collection light path and the waveguide.

Figure~\ref{fig:S3}(a) shows the simulated transmission as a function of the auxiliary beam waist position $(x,\ z)$, calculated using the angular spectrum method described in the main text. In general, the transmission decreases when the beam passes across the waveguide, because the larger divergence angle of the transmitted light prevents the collection objective from fully collecting the beam. When $x<0$, the transmission exhibits a small peak when the beam is precisely centered on the waveguide. 

Figure~\ref{fig:S3}(b) presents the experimentally measured relative transmission as a function of $z$ (red triangles). Values greater than unity arise because the relative transmission is obtained by directly dividing the photoelectric detector readings after and before passing across the waveguide. The black solid line is the simulated transmission for the waveguide distance $x = -5\,\mathrm{\mu m}$, which best matches the experimental data, selected by least-squares fitting. The gray shaded region shows the envelope bounded by the simulations for the $x = -5\pm1\,\mathrm{\mu m}$. To compare with the experiment, all simulated transmission curves in Fig.~\ref{fig:S3}(b) at different $x$ are normalized before fitting such that their value at $z=0\;\mu$m coincides with the measured relative transmission at $z=0\;\mu$m.  The experimental data agree well with the simulation, both showing a deep transmission dip near the waveguide position and a small peak at the center. The slight asymmetry of the experimental data is likely caused by the collection path and the incident path being slightly non-collinear.

Figure~\ref{fig:S3}(c) shows the alignment between the fluorescence collection path and the auxiliary beam. The blue squares and red circles represent the centers of the back-injected spot of the collection path and the auxiliary beam spot, respectively, obtained from repeated measurements. Due to slight camera jitter, the measured spot centers vary across measurements. Fitting the distribution of the centers with a normal distribution yields a 95.6\% confidence level that the distance between the two spot centers is less than $0.3\,\mathrm{\mu m}$. The blue and red dashed ellipses in the figure show the $2\sigma$ regions obtained from the Gaussian fits.

We performed the experiment described in Fig.~3 of the main text at various $z$ positions. To characterize the focusing effect of the collection light path, we fit the relationship between the counts ratio and $\Delta$ with an exponential function $\exp(-\Delta/\xi)$. The resulting $\xi$ as a function of $z$ is shown by the blue circles in Fig.~\ref{fig:S3}(d). The dashed line and the gray shaded region represent the reference measurement and its uncertainty, respectively, obtained from control experiments without the waveguide. The results indicate that the nanolensing effect is present only when the alignment error between the collection path and the waveguide is less than $2\,\mathrm{\mu m}$.

\section{Appendix D: Influence of the conveyor beams}

\begin{figure}[h]
\centering
\includegraphics[width=1.0\textwidth]{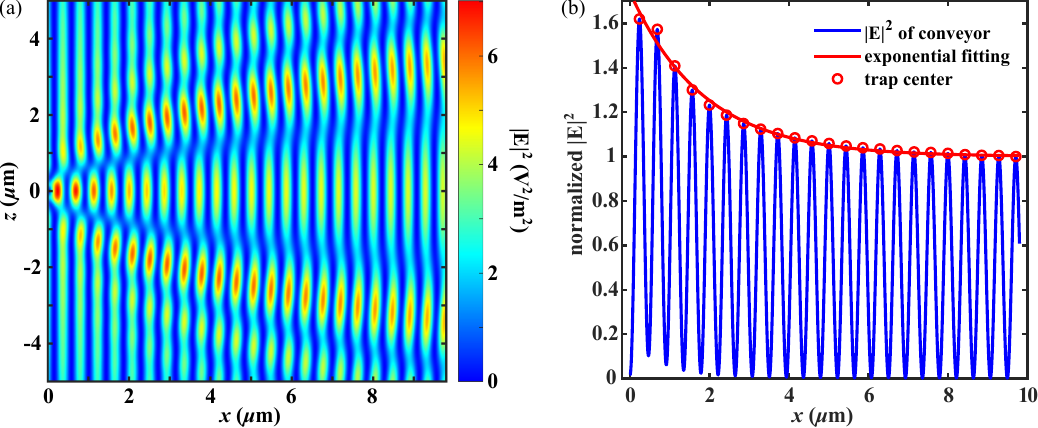}
\caption{The optical conveyor modulated by the nanolensing effect. (a) Numerical simulated $|\mathbf{E}(x)|^2$ distribution of the conveyor beams modulated by the nanolensing effect. (b) The blue line is the normalized $|\mathbf{E}(x)|^2$ distribution of the conveyor beams along the $z=0$ transversal line, and red circles represent the position of the peaks. The red line is an exponential fitting of the peaks, that is $|\mathbf{E}_{trap}(x)|^2$. The $|\mathbf{E}(x)|^2$ is normalized to the peaks $10\,\mathrm{\mu m}$ away from the waveguide, which is equivalent to the condition without the nanolensing effect.}
\label{figS4}
\end{figure}

In Fig. 3(b), the blue line is a corrected result, considering the nanolensing effect on the conveyor beams. 
In this section, we will discuss this in detail.
Fig.~\ref{figS4}(a) is the numerical simulated $|\mathbf{E}(x)|^2$ distribution of the conveyor beams modulated by the nanolensing effect.
The $x$ and $z$ axes are the same as Fig. 3(a). 
There are two possible effects due to the conveyor beams.
First, the closer the standing wave trap is to the waveguide, the deeper the trap, and the atoms in the trap will perceive a larger AC-Stark shift. 
This causes the atoms in each trap to be excited unevenly by the probe beam, which could be represented by an exciting factor $\theta(x)$.
Second, the closer the standing wave trap is to the waveguide, the more compact in the z direction it is, resulting in a smaller range of atomic motion.
This may cause the atomic fluorescence collection efficiency $\eta(x)$ to be subject to a correction $\zeta(x)$.
Thus, the expression 1 turns to:

\begin{equation}
    C(\Delta)=\frac{\int_{\Delta}^\infty \eta(x)\theta(x)\zeta(x)\mathrm{d}x}{\int_{0}^\infty \eta(x)\theta(x)\zeta(x)\mathrm{d}x}
\label{CR1_1}
\end{equation}

$\theta(x)$ is the probability of the atom being excited to the upper level, which is $\frac{I}{I_{sat}}\cdot\frac{\Gamma^{2}}{8}/[{\delta(x)}^{2}+\frac{\Gamma^{2}}{4}\left(1+\frac{I}{I_{sat}}\right)]$. 
$I$ is the probe light intensity. 
$I_{sat}$ is the atomic saturation light intensity. 
$\Gamma$ is the atomic spontaneous emission rate. 
$\delta(x)$ is the light detuning, which consists of the detuning of the probe beam $-2\pi\times28\,\mathrm{MHz}$ and the AC-Stark shift caused by the conveyor beams ($\approx 1.08 U(x)/\hbar$). 
$U(x)$ is the potential of the standing wave traps, which can be derived from Fig.~\ref{figS4}, and 1.08 is the ratio of the cyclic transition frequency shift to the ground-state energy level shift.
The blue line in Fig.~\ref{figS4}(b) is the normalized $|\mathbf{E}(x)|^2$ distribution of the conveyor beams along the $z=0$ transversal line, and red circles represent the position of the peaks. The red line is an exponential fitting of the peaks, that is $|\mathbf{E}_{trap}(x)|^2$. 
The $|\mathbf{E}(x)|^2$ is normalized to the peaks $10\,\mathrm{\mu m}$ away from the waveguide, which is equivalent to the condition without the nanolensing effect.
Since $U(x)\propto-|\mathbf{E}_{trap}(x)|^2$, and the standing wave trap depth without nanolensing effect is $\mathrm{k_B}\times1.25\,\mathrm{mK}$, $\theta(x)$ can be easily obtained.

$\zeta(x)$ can be derived from $\int \exp\left(-\frac{2z^2}{w_c^2(x)}\right) \frac{1}{\sqrt{2\pi}\sigma(x)} \exp\left(-\frac{z^2}{2\sigma^2(x)}\right) \, \mathrm{d}z$.
The first part in the integration $\exp\left(-\frac{2z^2}{w_c^2(x)}\right)$ is the fluorescence collection efficiency normalized to the $z=0$ situation, where $w_c(x)$ is the waist of the fluorescence collection light path at different $x$.
The second part $\frac{1}{\sqrt{2\pi}\sigma(x)} \exp\left(-\frac{z^2}{2\sigma^2(x)}\right)$ is the atomic number density distribution in the z direction in standing wave traps, where $\sigma(x)=w_t(x)\sqrt{\frac{k_BT}{4|U(x)|}}$.
$T$ is the atom temperature and $w_t(x)$ is the z-direction waist of standing wave traps at different $x$.
$\zeta(x)$ finally turns to $1/\sqrt{\frac{k_BTw_t^2(x)}{|U(x)|w_c^2(x)} + 1}$.
Since the fluorescence collection light path and the conveyor beams are modulated by the same waveguide, $\frac{Tw_t^2(x)}{|U(x)|w_c^2(x)}$ is  not going to change drastically.
Also, in our experiment system, $T\approx100\,\mathrm{\mu K}$ and $|U(x)|/\mathrm{k_B}>1.25\,\mathrm{mK}>10T$.
This makes $\zeta(x)$ always approximately equal to 1 and thus can be eliminated from expression~\ref{CR1_1}.

\section{Appendix E: Counts of the first 10 ms}
\begin{figure}[h]
\centering
\includegraphics[scale=0.88]{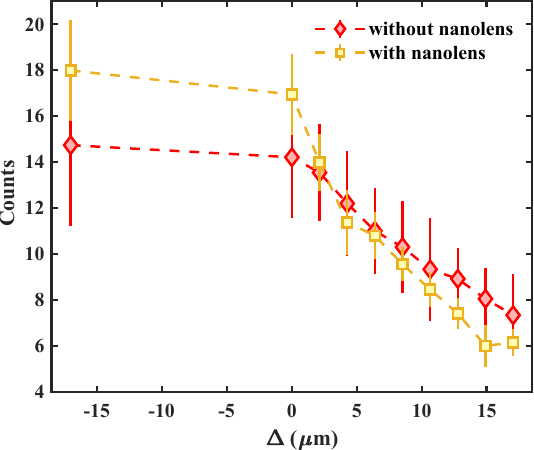}
\caption{Comparison of the first-$10\,\mathrm{ms}$ fluorescence counts with and without nanolens. The error bars represent the standard deviation of six experiments.}
\label{figS5}
\end{figure}

Comparison of the first-$10\,\mathrm{ms}$ fluorescence counts with and without nanolens is shown by Fig.~\ref{figS5}.
The data points of $\Delta = -17\,\mathrm{\mu m}$ indicate that the transport loss is almost zero.

\section{Appendix F: Single-atom signal criterion}
We take the integration time of the probe process as $40\,\mathrm{ms}$ and use the following three thresholds ($h_{1,2,3}$) to identify such single-atom events:

\begin{enumerate}
    \item Starting from the second data point after the probe beginning, the maximum count difference between the two data points before and after is greater than $h_1$, and the step is supposed to be located at the maximum.
    \item The count at the step minus the maximum count behind the step is greater than $h_2$.
    \item The count at the step minus the average count before the step is less than $h_3$.
\end{enumerate}

The first criterion is to distinguish the step signal of a single atom from the fluorescence fluctuation.
The second criterion is to ensure that the count after the step does not rise again, avoiding the case of accidental reduction of the count.
The third criterion is to ensure that the overall count before the step is at a higher level, avoiding the situation of an accidental increase in the count.
We set $h_1=2 \sqrt{2} \sigma_{bg}$, $h_2=\sigma_{bg}$, and $h_3=0.2  H_s$, where $\sigma_{bg}=8.46$ is the standard deviation of background counts obtained by a Gaussian fitting and $H_s$ is the height of the step signal itself.